\documentclass[12pt,preprint]{aastex}
\usepackage{graphics}
\usepackage{epsfig}
\usepackage{longtable}
\def\gsim{\lower 2pt \hbox{$\, \buildrel {\scriptstyle >}\over
{\scriptstyle \sim}\,$}}
\def\lsim{\lower 2pt \hbox{$\, \buildrel {\scriptstyle <}\over
{\scriptstyle \sim}\,$}}

\def\chandra{{\em Chandra\ }}

\def\cor{\widehat=}

\def\xs{M31}
\shortauthors{}
\shorttitle{}

\begin{document}

\title{{\sl Chandra} detection of diffuse hot gas in and around the \xs\ bulge}
\author{Zhiyuan Li, \& Q. Daniel Wang}
\affil{Department of Astronomy, University of Massachusetts \\
710 North Pleasant Street, Amherst, MA 01003, U.S.A. \\
 Email: zyli@nova.astro.umass.edu, wqd@astro.umass.edu}

\begin{abstract}
We report the detection of diffuse hot gas in \xs, using archival {\sl Chandra} observations
which allow us to map out a 30$^\prime \times 30^\prime$ field (covering
a galactocentric radius up to 4.5 kpc) and to detect sources in the
galaxy down to a 0.5-8 keV luminosity limit of $\sim10^{35}{\rm~ergs~s^{-1}}$. We
estimate the remaining stellar contribution from fainter
X-ray sources (primarily cataclysmic variables and coronally active binaries),
assuming that they spatially follow the stellar distribution. Indeed, the near-IR K-band
light of the galaxy closely traces the 2-8 keV unresolved X-rays, indicating 
a collective stellar X-ray emissivity consistent with those determined for the Galactic ridge and
M32, whereas the amount of the 0.5-2 keV unresolved emission is significantly greater than the expected stellar contribution, 
especially within a galactocentric radius of $\sim$2 kpc. 
Morphologically, this soft X-ray excess appears
substantially rounder than the bulge as seen in K-band and is elongated approximately
along the minor-axis at large radii. The excess thus most likely represents
the emission of diffuse hot gas in and around the galactic bulge.
Furthermore, the near side of the \xs\ disk casts an apparent shadow against
the soft X-ray excess, indicating that the hot gas extends to at least 2.5 kpc from
the galactic plane. We briefly discuss the implications of these results on 
the energy balance in the \xs\ bulge and on understanding the large-scale 
soft X-ray enhancement observed toward the
inner region of our own Galaxy.
\end{abstract}

\section{Introduction} {\label{sec:intro}}

The bulge of a galaxy, though consisting of mainly old stars, is a mecca of high-energy
activities. Low-mass X-ray binaries (LMXBs), in the luminosity range of $\sim 10^{35} -
10^{38} {\rm~ergs~s^{-1}}$, are among the brightest X-ray sources observed in 
a galaxy like our own. At lower luminosities, typically $\sim 10^{30} - 10^{33} 
{\rm~ergs~s^{-1}}$, are numorous cataclysmic variables (CVs) and coronally active 
binaries (ABs) that can be individually detected normally only in the Solar neighborhood.
Such stars are most likely responsible for the bulk of the unresolved 2-10 keV emission 
observed in the Galactic bulge/ridge (Revnivtsev et al.~2006), whereas lower energy X-rays 
from the same regions are subject to heavy interstellar absorption and hence difficult to
detect. One thus needs external perspectives of nearby galaxies. 
Indeed, Revnivtsev et al.~(2007) 
have shown that the unresolved emission from the low-mass bulge-dominated galaxy 
M32 over the entire 0.5-7 keV range is primarily stellar in origin. 

The interstellar 
medium (ISM) in a galactic bulge is also expected to be extremely energetic, 
chiefly due to a concentration of Type Ia supernovae (SNe). The bulk of the
mechanical energy release from such SNe is expected to be in shock-heated gas which
can be naturally traced by its X-ray emission. However, the observed luminosity of 
the unresolved (source-removed) X-ray emission from a galactic bulge typically accounts
for only a small fraction (a few \%) of the expected SNe energy release. How the remaining
energy is dissipated remains unknown. It may be propagated into the large-scale 
halo of the host galaxy in a mechanical outflow or sound waves, for example. In any case, 
determining the fate of this ``missing'' stellar feedback energy is fundamentally important in our
understanding of its role in galaxy evolution (e.g., Wang 2007). 
Here we present the first step of such a study.

We detect the truly diffuse soft X-ray emission from the bulge of \xs\ --- 
the nearest spiral galaxy ($d\sim$780 kpc; 
1$^\prime\cor$0.23 kpc) that is similar to our own Galaxy.
The galaxy contains no AGN, so the non-nuclear X-ray emission can be studied even in
the very central region. The \xs\ bulge has little cool gas and 
star formation, so the X-ray contribution from young stellar populations is minimal. 
The moderate inclination ($\sim78^\circ$) of the \xs
disk and the relatively low Galactic foreground absorption (${\rm N_H }
\sim 6.7{\times}10^{20}{\rm~cm^{-2}}$) also allow us to 
detect extraplanar X-ray emission in the 0.5-2 keV range. Indeed,
detections of diffuse soft X-ray emission have been claimed, based on 
spectral decompositions (Shirey et al.~2001; Takahashi et al.~2004).
However, such a decomposition depends 
sensitively on rather arbitrary choices of spectral models for various components contributing to the spectrum, 
both diffuse and discrete. Our approach here is to spatially map out the
diffuse X-ray emission (in addition to its energy dependence) and to study its 
relationship to other stellar and interstellar components of the galaxy.
While details of the data reduction as well as a comprehensive analysis of 
the sources and modeling of the hot gas will be presented elsewhere 
(Li et al.~2007, in preparation), we focus here in presenting the
detection of the diffuse hot gas in and around the \xs\ bulge and its
immediate implications. We quote our statistical error estimates at the 90\% confidence level.

\section{Data preparation} {\label{sec:data}}

Our X-ray study was based on 31 \chandra ACIS archival observations of \xs\ taken by 2005. 
The majority (21 out of 31) of these observations were taken with the ACIS-I array
and aimed toward the \xs\ bulge with the aim-points located within 1$^\prime$ from the galactic center. To maximize the coverage and uniformity of the combined field, 
we utilized data only from the front-illuminated CCDs (the ACIS-I array and the S2 
chips) of the 21 observations. For same reason, we also included I-chip data from four 
ACIS-S observations. These data together cover a field of $r\sim$18$^\prime$ around
the center of \xs. Furthermore, for local sky background determination, we 
used six additional ACIS-I observations which were aimed toward an ``off-field'' 
$\sim$$20^\prime$ southwest to the center. 

We reprocessed the data using CIAO (version 3.3), following the {\sl Chandra} ACIS data analysis guide. 
We generated count and exposure maps for each observation in the 0.5-1, 1-2, 2-4 and
4-8 keV bands. Corresponding instrumental background maps were generated from the ``stowed'' data, after calibrating
the 10-12 keV count rate with individual observations. 
The total effective exposure is $\sim$95 ks in the central region and gradually drops to $\lesssim$ 20 ks at 
radii $r \gtrsim 10^\prime$. 

\begin{figure*}[!htb]
  \centerline{
       \epsfig{figure=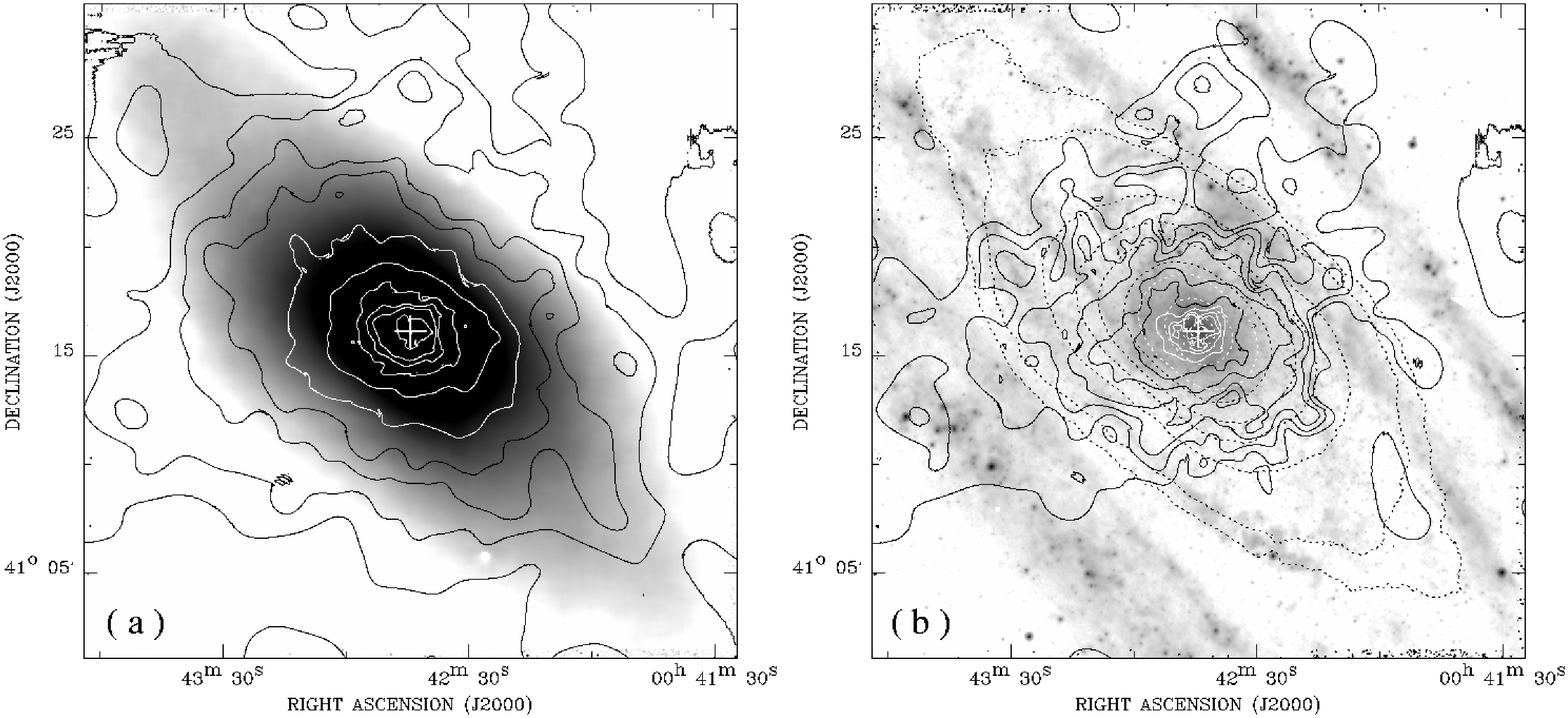,width=1.0\textwidth,angle=0}
     }
  \caption{(a) Background-subtracted, exposure-corrected and smoothed intensity contours of the {\sl Chandra} ACIS-I 0.5-2 keV unresolved emission overlaid on the 2MASS K-band image of \xs. The contours are at 
3, 6, 10, 16, 32, 64, 128, 196, 256 and 512 ${\times}10^{-4}{\rm~cts~s^{-1}~arcmin^{-2}}$. 
The galactic center is marked by a plus sign.
(b) Contours of the diffuse (stellar contribution-subtracted) 
X-ray intensity (solid) and K-band light (dotted) overlaid on the 
{\sl Spitzer} MIPS 24 $\mu$m image.
}
 \label{fig:unr}
\end{figure*}

Following a procedure detailed in Wang (2004), we performed source detection 
in the soft (0.5-2 keV), hard (2-8 keV) and broad (0.5-8 keV) bands. With a local false detection 
probability $P \leq 10^{-6}$, a total of 305 sources are detected in the field. 
To study the unresolved X-ray emission,
we excluded each of the sources from maps of individual observations with  circular regions enclosing $\sim$97\% of the source counts. The residual of this source removal 
contributes about 10\% of the remaining unresolved X-ray emission in the field.  
The source-removed maps were then reprojected to generate combined images in the four bands. We further statistically corrected for the variation of the detection incompleteness across the field, to a common detection limit of 8$\times10^{34}{\rm~ergs~s^{-1}}$ (0.5-8 keV). 
Because of the 
relatively flat luminosity function of the sources (mostly LMXBs; Li et al. 2007 in preparation; see also Voss \& Gilfanov 2007), 
the correction (normalized according to the 2MASS K-band intensity; Fig.~\ref{fig:unr}a; Jarrett et al.~2003) typically amounts to less than 6\% of the unresolved emission. For the same reason, the residual contribution 
from LMXBs at lower luminosities is found to be negligible. 

\section{Analysis and results} {\label{sec:unresolved}}

Fig.~\ref{fig:unr}a shows the 0.5-2 keV unresolved X-ray emission from a 30$^\prime$ by 30$^\prime$ region around the center of \xs, compared with the 
near-IR K-band image. 
In the inner bulge and along the major-axis of the disk, the X-ray emission shows morphological similarities with the K-band 
light, whereas at large radii the X-ray morphology appears substantially rounder and is elongated 
approximately along the minor-axis, in particular at the southeast side, 
indicating the presence of diffuse hot gas.

\subsection{The collective stellar emission} {\label{subsec:spat_anal}}
To quantify the diffuse hot gas, we need to isolate the collective stellar contribution,
which presumably spatially follows the K-band light distribution. Optimal for this purpose is to inspect regions
along the major-axis, where the soft X-ray emission morphologically mimics the K-band light better than in regions further away from the major-axis (Fig.~\ref{fig:unr}a).
Fig.~\ref{fig:rsb}a shows the unresolved X-ray intensity profiles along the major-axis, together with the corresponding K-band intensity profile.
Indeed, the K-band profile as a ``model''  fits the hard band profile 
(triangles in Fig.~\ref{fig:rsb}a)  well with a normalization factor 
$N_K = 4.3\pm0.2{\times}10^{-5}{\rm~cts~s^{-1}~arcmin^{-2}}/({\rm MJy~sr^{-1}})$. 
Therefore, the hard band X-ray emission is fully consistent with an origin in the old stellar population.

The collective stellar emission should also contribute at lower energies. 
However, while the soft X-ray profile (diamonds in Fig.~\ref{fig:rsb}a) 
can match that of the K-band light reasonably well at major-axis radii $\gtrsim 
8^\prime$, there is a clear excess above the collective stellar 
contribution in the inner region. This soft excess in the bulge
is another indication for the presence of hot gas, although its exact spatial
distribution is yet to be determined. We include an exponential law to
approximately account for the excess. The resultant fit is satisfactory 
(solid curve in Fig.~\ref{fig:rsb}a), with a fitted $N_K = (25.0\pm2.4) {\times}10^{-5}{\rm~cts~s^{-1}~arcmin^{-2}}/({\rm MJy~sr^{-1}})$. The fit predicts that the stellar component contributes about 60\% to the soft emission within a major-axis raidus of $\sim$8$^{\prime}$ and becomes dominant further beyond.  
The fit also reveals an interesting drop of the X-ray to K-band
intensity ratio within the central 0\farcm5 (more clearly indicated in a radial intensity profile which is not shown here),
with values still comparable to or greater than that at the large major-axis radii.
The nature of this ratio drop is currrently being investigated, and we note that it has little effect on the result of the above procedure.

\subsection{The diffuse X-ray emission} {\label{subsec:dif}}
Fig.~\ref{fig:unr}b shows an image of the truly diffuse emission after 
subtraction of the collective 
stellar contribution from the total unresolved X-ray emission of \xs.
The emission along the major-axis is confined within a projected distance 
of $\sim$$8^\prime$ ($\sim$1.8 kpc) from the galactic center to the southwest 
and is slightly more extended to the northeast. The overall morphology 
is elongated approximately along the minor-axis, with an extent of more than 15$^\prime$ 
($\sim$3.5 kpc) on both sides with respect to
the center; but the emission appears considerably fainter
on the northwest side and somewhat interrupted by the presence of 
spiral arms. This asymmetry is further illustrated in Fig.~\ref{fig:rsb}b, where diffuse X-ray intensity profiles along the minor-axis are shown separately in the 0.5-1 keV and 1-2 keV bands.  
The asymmetry can be naturally explained by the soft X-ray absorption 
of the galactic disk, as its near side is to the northwest. 
In particular, a major spiral arm and the star-forming ring, as traced by the peaks of the
{\sl Spitzer} MIPS 24 $\mu$m emission (Gordan et al.~2006), apparently cast deep X-ray shadows on the northwestern side. 
Estimated from the relative depth of these shadows, the equivalent X-ray-absorbing column densities are $\sim$$1.2{\times}10^{21}{\rm~cm^{-2}}$ and $\sim$$3.6{\times}10^{21}{\rm~cm^{-2}}$, consistent with the hydrogen column densities 
of the spiral arm and the star-forming ring (Nieten et al.~2006). 
No similar shadow is apparent on the southeastern side, indicating 
that the emission on this side is mostly from the bulge region in front of the disk. 
Therefore, the diffuse emission seems to have an intrinsic (absorption-corrected) 
overall coherent morphology reminiscent of a bi-polar outflow from the bulge.
The soft X-ray absorption by the disk suggests that the vertical 
extent of the X-ray-emitting gas from the galactic plane 
is at least 2.5 kpc. 

We characterize the profiles (Fig.~\ref{fig:rsb}b) 
at distances of $-7^\prime<z<0^\prime$ off the major-axis 
with an exponential law: $I(z)=I_g e^{-|z|/z_0}$, where $I_g$ is the central intensity
and $z_0$ is the projected scale-height. The best-fit $z_0$, being 2\farcm5$\pm$0\farcm1 ($\sim$0.6 kpc), shows no statistically significant difference between the two bands, indicating that the hot gas in the bulge has 
little temperature variation.
The hardness ratio, $I_{g,1-2 keV}/I_{g,0.5-1 keV}$$\sim$0.25, is consistent with a spectrum from an isothermal
gas with a temperature of $\sim$0.4 keV, subject to the Galactic foreground absorption. For comparison, the hardness ratio of the stellar emission is $\sim$0.72; 
the scale-heigh of the K-band light is $\sim$2\farcm1, if characterized by an exponential law as well. Therefore, the diffuse X-ray emission is both softer and more extended than
the stellar contribution.
At $|z| \gtrsim 7^\prime$, however, the soft X-ray intensity distribution levels off
(Fig.~\ref{fig:rsb}b). Part of this leveling may be related to emission 
associated with the galactic disk, partially compensating its X-ray absorption. The
exact intensity level at such large distances, however, depends on an accurate 
subtraction of the local sky background which has been estimated in the off-field  
and may be biased (\S~2). The sky coverage of the present ACIS observations 
is still too limited to accurately determine both the background and the large-scale distribution of the
diffuse emission associated with the bulge. 
Assuming the above exponential fit and temperature estimate as well as 
an intrinsic symmetry with respect to the galactic 
plane, we infer a 0.5-2 keV luminosity of the diffuse emission as  
$\sim$2.2$\times10^{38}{\rm~ergs~s^{-1}}$. The flat tail parts of the 
profiles give an additional $\sim$3$\times10^{37}{\rm~ergs~s^{-1}}$.

\section{Discussion} {\label{sec:discussion}}
In \S~\ref{subsec:spat_anal} we have estimated the stellar contribution to the total unresolved X-ray emission,
which is a key step in isolating the truly diffuse emission. It is thus instructive to 
compare our result with independent measurements.
Sazonov et al.~(2006) measured the collective X-ray emissivity (per unit stellar mass) of the old stellar populations in the Solar neighborhood
to be $9{\pm}3\times10^{27}{\rm~ergs~s^{-1}~M_{\odot}^{-1}}$
in the 0.5-2 keV band and $3.1{\pm}0.8\times10^{27}{\rm~ergs~s^{-1}~M_{\odot}^{-1}}$ in the 2-10 keV band. 
Revnivtsev et al.~(2006) showed that the Galactic ridge X-ray emission closely follows the near-IR light that traces the Galactic stellar mass distribution,
and that the X-ray to near-IR intensity ratio is consistent with the collective X-ray emissivity of old stellar populations inferred from the Solar neighborhood.
Revnivtsev et al.~(2007) further found that the 0.5-7 keV unresolved X-ray emission
and K-band stellar light in M32 have consistent spatial distributions,
but they did not explicitly give fitted parameters of the spectral model, 
which would allow for an immediate comparison with the \xs\ values. 

We have thus re-extracted the unresolved X-ray spectrum of M32
from two {\sl Chandra} ACIS-S observations (Obs.ID.~2017 and 5690) with a total exposure of 160 ks. The spectrum can be adequately 
fitted by a model consisting of a power-law component (with a photon index of 1.86$^{+0.26}_{-0.21}$) and a thermal
plasma component (temperature of 0.45$^{+0.15}_{-0.10}$ keV), with the Galactic foreground absorption. The collective emissivity is 
$5.8{\pm}1.1 (5.6{\pm}1.0)\times10^{27}{\rm~ergs~s^{-1}~M_{\odot}^{-1}}$ in the 0.5-2 (2-10) keV band, consistent with the values reported by Revnivtsev et al.~(2007). The spectral model-predicted ACIS-I count rate
is $\sim$17.1 (3.9) ${\times}10^{-5}{\rm~cts~s^{-1}~arcmin^{-2}}/({\rm MJy~sr^{-1}})$ in the 0.5-2 (2-8) keV band.
The hard band value agrees well with the measurement for \xs\ (4.3$\pm0.2 {\times}10^{-5}$), 
accounting for the residual photons spilling outside the source-removal regions ($\sim 10\%$; \S~\ref{sec:data}). The soft band stellar emissivity of \xs\ (residual LMXB contribution excluded) is a factor of $1.4\pm0.3$ higher than the M32 value, but 
is consistent with that inferred from the Solar neighborhood. This 
discrepancy
is not totally unexpected, considering various statistical errors and 
hiden systematic uncertainties (e.g., in the mass-to-light ratio and in 
the spatial and spectral modeling). We note that adopting 
the M32 stellar emissivity to remove the stellar contribution
in \xs\ would enhance the hot gas contribution in and around the \xs\ bulge, 
but would not qualitatively alter the picture of the diffuse emission 
presented in \S~\ref{subsec:dif} and below. 
   

The characterization of the diffuse hot gas sheds important insights 
into the energy balance in the \xs\ bulge.
The estimated luminosity of the hot gas ($2.5\times10^{38}{\rm~ergs~s^{-1}}$) is only about 0.6\% of the expected SNe mechanical energy input, 
$\sim$$4\times10^{40}{\rm~ergs~s^{-1}}$. As mentioned in \S~\ref{sec:intro}, 
this indicates that the input energy may be removed primarily in an outflow.
Dynamically, such an outflow tends to find 
its way along steeper pressure gradient against the gravity of the galaxy,
consistent with the observed bi-polar morphology of the diffuse X-ray emission.
If the gas were quasi-static, one would expect its distribution to follow 
that of the gravitational potential, i.e., more extended along the major-axis. 
However, the gas may not
be hot enough to ultimately escape from the deep gravitational potential of \xs;
it is also not clear how the outflow interacts with the large-scale halo
of \xs\ and how the mechanical energy is dissipated.  
Similar considerations also challenge the studies of 
X-ray-faint elliptical galaxies (e.g., David et al.~2006).
Ongoing numerical simulations would help to understand the nature of the hot gas and its role in the evolution of these systems.  

\begin{figure*}[!htb]
  \centerline{
      \epsfig{figure=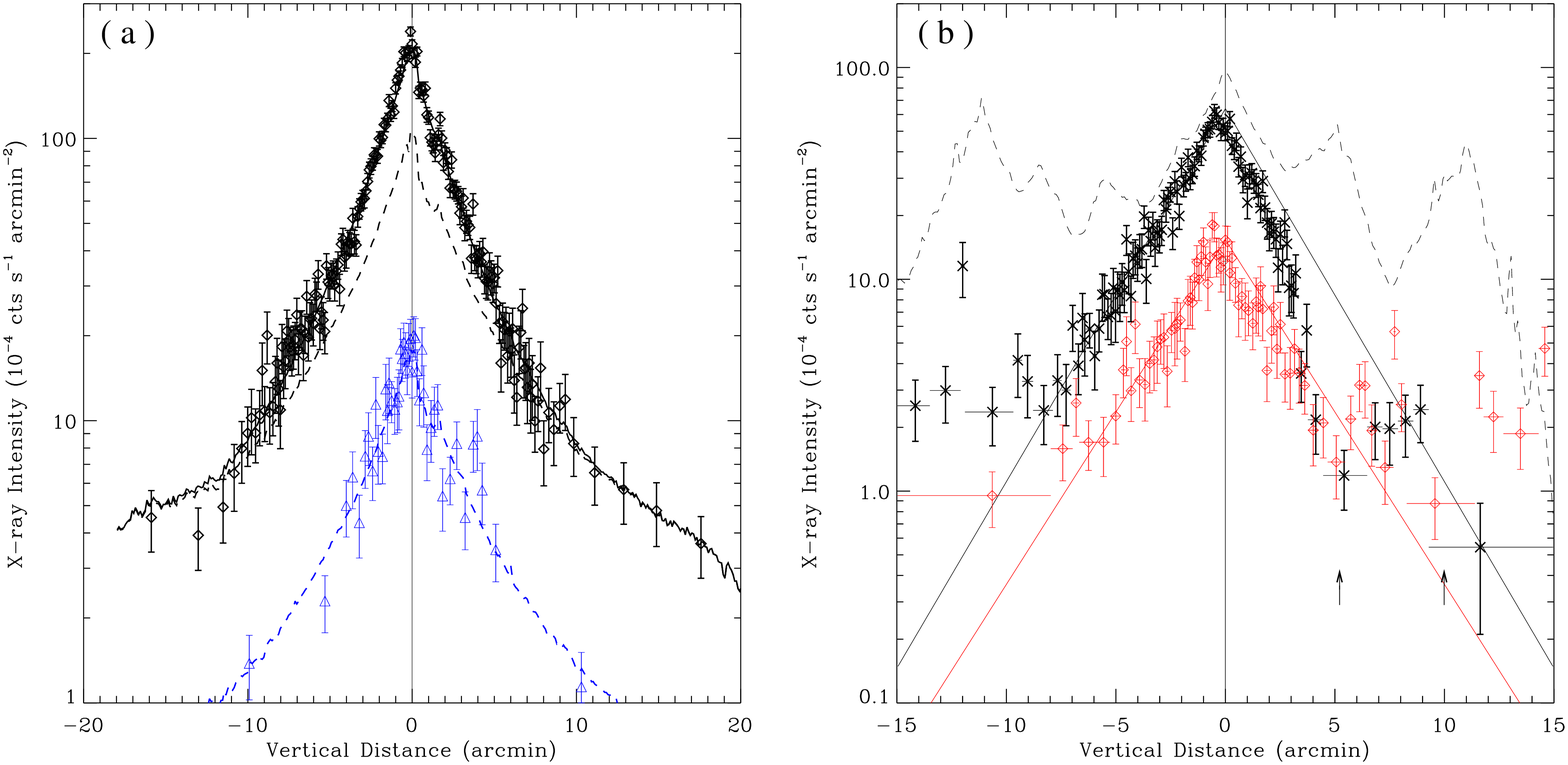,width=1.\textwidth,angle=0}
    }
  \caption{(a) 0.5-2 ({\sl diamonds}) and 2-8 keV ({\sl triangles}) keV intensity profiles of the unresolved X-ray emission along the major-axis.
A position angle of 45$^\circ$ is adopted.
The positive side is toward the southwest. 
The full width for averaging the intensity is 8$^\prime$. 
Spatial binning is adaptively ajusted to achieve a signal-to-noise ratio
better than 4, with a minimum step of 6$^{\prime\prime}$. 
The X-ray profiles are characterized by a normalized K-band intensity profile (dashed curves), 
and an additional exponential law for the soft band (solid curve).
(b) 0.5-1 ({\sl crosses}) and 1-2 keV ({\sl diamonds}) intensity profiles of the diffuse emission along the minor-axis; the stellar contribution
has been subtracted.
The positive side is toward the northwest. The full width for averaging the intensity is 16$^\prime$. 
The adaptive steps achieve a signal-to-noise ratio better than 3. 
The solid curves represent a fit with an exponential law.
The corresponding 24 ${\mu}$m intensity profile is shown by the dashed curve. 
The arrows mark the positions of the shadows casted by a spiral arm and the star-forming ring. 
}
 \label{fig:rsb}
\end{figure*}

Our unambiguous detection of the diffuse hot gas in 
and around the \xs\ bulge also helps to understand the soft X-ray enhancement observed toward the inner
region of our Galaxy. The temperature of the hot gas associated with the 
\xs\ bulge, 0.4 keV, is similar to that with the Galactic bulge, as estimated
from the ROSAT all-sky survey (Snowden et al.~1997). Based on 
a hydrostatic model of the Galactic bulge X-ray emission developed by Wang (1997), Almy 
et al.~(2000) further inferred a total 0.5-2 keV luminosity of $\sim8 \times 
10^{38}{\rm~ergs~s^{-1}}$, about four times greater than 
our estimated \xs\ bulge luminosity. The relatively high luminosity of the Galactic bulge
manifests in the large extent of the soft X-ray enhancement from the Galactic 
bulge. At Galactic lattitudes $b\sim-15^\circ$ ($\sim$2 kpc from the plane), for example, where both
the confusion with the foreground emission features and the interstellar absorption are
relatively small, the intensity has an averaged value of $\sim$$6 (4){\times}10^{-4} {\rm~ROSAT~PSPC~cts~s^{-1}~arcmin^{-2}}$ in the 0.75 (1.5) keV band (Snowden et al.~1997). Had this emission been detected from \xs\ by the {\sl Chandra}
ACIS-I, it would be measured with an intensity of $\sim$$10 (3){\times}10^{-4}{\rm~cts~s^{-1}~arcmin^{-2}}$ in the 0.5-1 (1-2) keV band, about 2-4 times higher than the observed \xs\ values represented by the tails (Fig.~\ref{fig:rsb}b). The intensity drops slowly and even shows local enhancements at high 
latitudes ($b\gtrsim-30^\circ$; Snowden et al. 1997). Within $|b| \lesssim 10^\circ$,
the interstellar absorption is severe, little can be inferred reliably about the properties
of the hot gas. It is in this corresponding region in the \xs\ bulge ($|z| \lesssim 6^\prime$)
that the diffuse soft X-ray intensity shows the steepest increase 
(by about one order of magnitute) toward the galactic center. 
Such a mid-plane concentration of 
diffuse soft X-ray emission may also be present instrinsically in our 
Galactic bulge. Clealy, a more careful comparison and modeling 
of the X-ray data sets are needed in order to
understand the similarity and difference in the hot gas characteristics 
and their relationship to other galactic properties 
(e.g., the effect of recent active star formation in the Galactic center).

\acknowledgements
We thank D. Calzetti, M. Fardal, 
and S. Tang for 
helpful comments and discussions. This work is supported by the SAO 
grant AR7-8006X.

\begin{references}
\reference{} Almy, R. C., McCammon, D., Digel, S. W., Bronfman, L., \& May, J. 2000, ApJ, 545, 290 
\reference{} David, L. P., Jones, C., Forman, W., Vargas, I. M., \& Nulsen, P. 2006, ApJ, 653, 207
\reference{} Gordon, K. D., et al. 2006, ApJ, 638, L87
\reference{} Jarrett, T. H., Chester, T., Cutri R., Schneider, S. E., \& Huchra, J. P., 2003, \aj, 125, 525
\reference{} Nieten, Ch., et al.
2006, A\&A, 453, 459
\reference{} Revnivtsev, M., Sazonov, S., Gilfanov, M., Churazov, E., \& Sunyaev, R. 2006, A\&A, 452, 169
\reference{} Revnivtsev, M., Churazov, E., Sazonov, S., Forman, W., \& Jones, C. 2007, astro-ph/0702578
\reference{} Sazonov, S., Revnivtsev, M., Gilfanov, M., Churazov, E., \& Sunyaev, R. 2006, A\&A, 450, 117
\reference{} Shirey, R., et al. 2001, A\&A, 365, L195
\reference{} Snowden, S. L., et al. 1997, ApJ, 485, 125
\reference{} Takahashi, H., Okada, Y., Kokubun, M., \& Makishima, K. 2004, ApJ, 615, 242
\reference{} Voss, R., \& Gilfanov, M. 2007, A\&A, 468, 49
\reference{} Wang, Q. D. 1997, in IAU Colloq. 166, The Local Bubble and Beyond, ed. D. Breitschwerdt, M. Freyberg, \& J. Tr$\ddot{u}$mper (Springer: Berlin), 503
\reference{} Wang, Q. D. 2004, ApJ, 612, 159
\reference{} Wang, Q. D. 2007, EAS Publications Series, 24, 59
\end{references}
\end{document}